\documentclass[runningheads]{llncs}
\usepackage[T1]{fontenc}
\usepackage{graphicx,verbatim}
\usepackage{amssymb}
\usepackage{booktabs}
\usepackage{multirow}
\usepackage[table]{xcolor}
\usepackage{makecell}
\usepackage{threeparttable}
\usepackage{booktabs}
\usepackage{multirow}
\usepackage{colortbl}
\usepackage{tabularx}
\usepackage{amsmath}
\usepackage{threeparttable}
\usepackage{amsfonts}
\usepackage{array}
\usepackage{float}
\usepackage{ragged2e}  
\usepackage{hyperref}
\usepackage{adjustbox}
\begin{document}
\title{Revisiting Global Token Mixing in Task-Dependent MRI Restoration: Insights from Minimal Gated CNN Baselines}
\titlerunning{Revisiting Token Mixing for MRI Restoration}
%

\author{Xiangjian Hou\inst{1,2} \and
Chao Qin\inst{3} \and
Chang Ni\inst{1,4} \and
Xin Wang\inst{5} \and
Chun Yuan\inst{1} \and
Xiaodong Ma\inst{1}}
\authorrunning{X. Hou et al.}
%
\institute{Dept. of Radiology \& Imaging Sciences, University of Utah, Salt Lake City, UT, USA\\
\email{\{chun.yuan, xiaodong.ma\}@hsc.utah.edu, \{xiangjian.hou, chang.ni\}@utah.edu} \and
Dept. of Electrical \& Computer Engineering, University of Utah, Salt Lake City, UT, USA\\ \and
Dept. of Computer Vision, Mohamed bin Zayed University of Artificial Intelligence, Abu Dhabi, UAE\\
\email{chao.qin@mbzuai.ac.ae} \and
Dept. of Biomedical Engineering, University of Utah, Salt Lake City, UT, USA\\ \and
Dept. of Electrical \& Computer Engineering, University of Washington, Seattle, WA, USA\\
\email{xwang99@uw.edu}}

  
\maketitle              
\begin{abstract}
Global token mixing—implemented via self-attention or state-space sequence models—has become a popular model design choice for MRI restoration. However, MRI restoration tasks differ substantially in how their degradations vary over image and k-space domains, and in the degree to which global coupling is already imposed by physics-driven data consistency terms. In this work, we ask the question whether global token mixing is actually beneficial in each individual task across three representative settings: accelerated MRI reconstruction with explicit data consistency, MRI super-resolution with k-space center cropping, and denoising of clinical carotid MRI data with spatially heteroscedastic noise. To reduce confounding factors, we establish a controlled testbed comparing a minimal local gated CNN and its large-field variant, benchmarking them directly against state-of-the-art global models under aligned training and evaluation protocols. For accelerated MRI reconstruction, the minimal unrolled gated-CNN baseline is already highly competitive compared to recent token-mixing approaches in public reconstruction benchmarks, suggesting limited additional benefits when the forward model and data-consistency steps provide strong global constraints. For super-resolution, where low-frequency k-space data are largely preserved by the controlled low-pass degradation, local gated models remain competitive, and a lightweight large-field variant yields only modest improvements. In contrast, for denoising with pronounced spatially heteroscedastic noise, token-mixing models achieve the strongest overall performance, consistent with the need to estimate spatially varying reliability. In conclusion, our results demonstrate that the utility of global token mixing in MRI restoration is task-dependent, and it should be tailored to the underlying imaging physics and degradation structure. Code and pretrained models will be publicly released.


\end{abstract}

\section{Introduction}

Global token mixing has become a popular model design choice in image restoration, driven by transformer-based long-range interaction and more recent attention-free state-space mixers \cite{liang2021swinir,zamir2022restormer,guo2025mambair}.
Motivated by these advances, MRI restoration pipelines increasingly incorporate such global mixers in both reconstruction and enhancement models \cite{guo2023reconformer,fabian2022humus,huang2024mambamir,meng2025dh}.

However, the need for explicit global mixing in MRI is not self-evident.
In accelerated reconstruction, MRI inverse problems already embed global coupling through Fourier encoding and repeated physics-based data consistency steps in unrolled schemes~\cite{zbontar2018fastmri,sriram2020end,fabian2022humus}.
Moreover, degradation structures vary substantially across MRI restoration tasks: in k-space center-crop super-resolution, the forward model deterministically truncates high spatial frequencies, which can be viewed as a convolution with a sinc point-spread function;
thus restoration primarily requires recovering/injecting missing high-frequency details rather than re-inferring global anatomy~\cite{deng2024exploring}.
In contrast, denoising of clinical MRI data can exhibit strong spatial heteroscedasticity due to coil configuration and anatomy \cite{zeng2025deep}. These differences across tasks suggest that the benefit of global mixing may be task-dependent, rather than universally positive.
Meanwhile, current MRI restoration studies largely introduce new architectures under task-specific pipelines, and it remains unclear when global token mixing is truly necessary under protocol-aligned comparisons.

Motivated by these observations, we ask a focused question: When does global token mixing meaningfully help MRI restoration, and when is it less critical given physics and degradation structure?

To answer this question, we conduct what is, to the best of our knowledge, the first protocol aligned study investigating global token mixing across three MRI restoration settings: accelerated reconstruction, super resolution, and denoising. We use a minimal NAFNet style gated CNN as the shared backbone~\cite{chen2022simple}, and a lightweight large field extension obtained by adapting the LSConv~\cite{wang2025lsnet} to a dynamic gating block as a middle state of token mixing. 
Our results reveal that the utility of global token mixing is task dependent: it provides gains when degradations are spatially non uniform but becomes less critical when acquisition physics and data consistency steps already impose strong global coupling. 
These findings can guide the physics-tailored design of future MRI restoration models.


\subsection{Related Work}

Recent restoration backbones explore explicit mechanisms to exchange information across spatial locations.
Transformer-based restorers employ self-attention to model long-range interactions in restoration \cite{liang2021swinir,zhang2024xformerhybridxshapedtransformer}.
Recently, attention-free mixers have also been adapted to restoration, including state-space models (e.g., MambaIR) and RWKV-style restorers \cite{guo2025mambair,yang2025restore}. These mixer families have also been increasingly incorporated into MRI restoration pipelines for reconstruction and enhancement \cite{guo2023reconformer,meng2025dh,fabian2022humus}.
MambaOut~\cite{yu2025mambaout} revisits mamba-style mixers for vision and empirically suggests that the benefit of token mixer varies across tasks.

Meanwhile, convolutional restoration backbones still remain widely used; NAFNet proposes an activation-free with minimal gated block design built on a lightweight multiplicative gate~\cite{chen2022simple}.
Beyond purely local, fixed-kernel mixing, high-level recognition backbones have also explored context-conditioned convolution to enlarge the effective receptive field while keeping convolutional computation.
We adopt a Large-Small Convolution (LSConv)~\cite{wang2025lsnet}, where a large-kernel perception branch parameterizes location-wise small-kernel aggregation to the gated block. This design strikes an elegant balance between local CNN filtering and dense global token mixing, effectively expanding the receptive field without introducing costly all-to-all interactions.

MRI restoration spans distinct degradations and pipelines: accelerated reconstruction is typically studied with unrolled/cascaded frameworks that alternate explicit data consistency and a learned regularizer \cite{schlemper2017deep,sriram2020end,fabian2022humus,wang2024progressive}, while k-space center-crop super-resolution and dedicated-coil-absence denoising are often formulated as image-to-image restoration with controlled low-pass truncation or spatially varying noise \cite{deng2024exploring,zeng2025deep}.
Recent MRI models increasingly incorporate transformer/SSM-style mixers, especially in reconstruction \cite{guo2023reconformer,huang2024mambamir,meng2025dh,zou2025mmr}, yet it remains unclear whether such global mixing yields consistent benefits across tasks with different physics-imposed coupling and degradation structures.
We therefore study token mixing in a task-dependent manner across reconstruction, super-resolution, and denoising under aligned training and evaluation settings.

\section{Method}

We study three representative MRI restoration settings that differ in (i) the extent to which acquisition physics already enforces global coupling and (ii) how degradations vary across space (i.e., image domain) and frequency (i.e., k-space domain). In the following, we first formalize each task under a unified notation (Sec.~\ref{sec:problem_formulation}), and then describe our minimal gated-CNN baseline, its large-field extension, and the aligned protocols used to benchmark against state-of-the-art global models.

\subsection{Problem Formulation}\label{sec:problem_formulation}

\begin{figure}[!htbp]
\includegraphics[width=\textwidth]{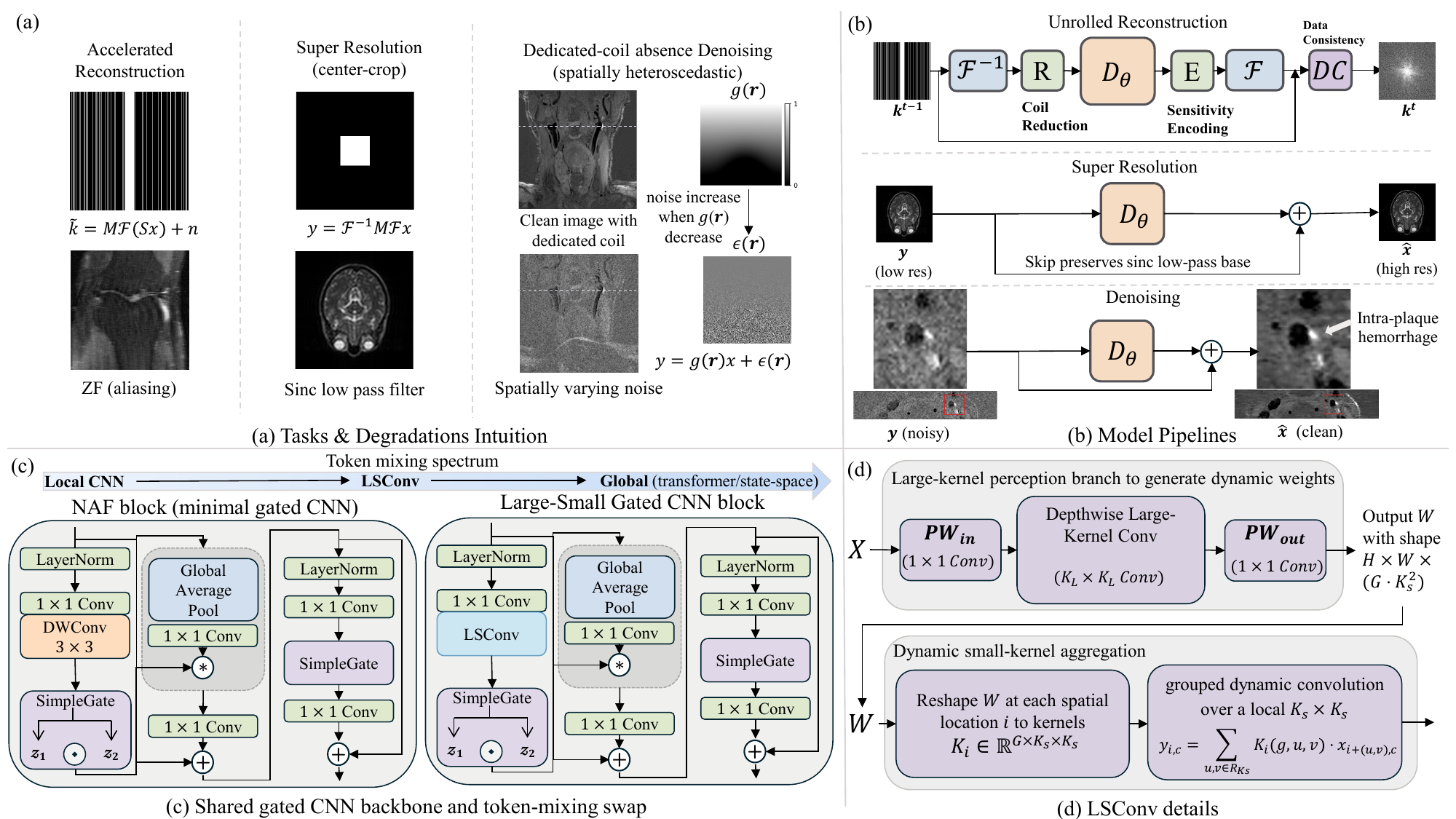}
\caption{Task-dependent MRI restoration framework. (a-b) Degradations and restoration pipelines across three tasks. (c-d) Shared CNN backbone comparing local and LSConv token mixing.} \label{fig1}
\end{figure}

\textbf{Accelerated Reconstruction}
For a target image $x \in \mathbb{C}^{N}$ and coil sensitivity maps $S$, the undersampled multi-coil $k$-space data are given by $\tilde{k} = M \mathcal{F}(Sx) + n$, where $\mathcal{F}$ is the Fourier transform and $M$ is the sampling mask.
We solve the inverse problem using an unrolled scheme that alternates explicit data consistency with a learned image-domain correction $D_\theta$.

Let $E$ and $R$ denote the sensitivity encoding and reduction (adjoint) operators, respectively. At iteration $t$, we update the $k$-space estimate $k^t$ via:
\begin{equation}
k^{t+1}
=
k^{t}
-
\mu_t\, M\bigl(k^{t} - \tilde{k}\bigr)
+
\mathcal{F} E \, D_\theta\!\bigl(R \mathcal{F}^{-1}(k^t)\bigr)
\label{eq:ar_unroll}
\end{equation}
where the middle term enforces data consistency on sampled locations, and the final term injects the learned regularizer.
After $T$ (eight) iterations, the final image is recovered as $\hat{x} = R \mathcal{F}^{-1}(k^{T})$. 
Because global coupling is already imposed by $\mathcal{F}$ and repeated data-consistency steps, we expect limited additional benefit from global token mixing inside $D_\theta$—a hypothesis we will test in Sec.~\ref{sec.result}.

\textbf{Setup}
We evaluate on FastMRI Knee~\cite{zbontar2018fastmri} (973/199 train/val scans) and Stanford 2D FSE~\cite{cheng2018stanford2dfse} (89 volumes, 80/20 split), following previous studies~\cite{wang2024progressive,fabian2022humus}.
Single-coil images are cropped to $320^2$, Fourier transformed to k-space, and then masked at $4\times$ (8\% center fullsampled) or $8\times$ (4\% center fullsampled).
Multi-coil data ($C=15$) are cropped to $384^2$ with $8\times$ acceleration (4\% center).
For multi-coil, we instantiate our model within the sensitivity-aware framework; to ensure fair comparison, we benchmark against methods natively operating within this physics-driven paradigm.

\textbf{Super-Resolution}
MRI Super-Resolution (SR) via \emph{k}-space center cropping $y = \mathcal{F}^{-1} M \mathcal{F} x$ constitutes an ideal low-pass degradation. Following~\cite{deng2024exploring}, SR networks decouple into a linear low-pass filter and a non-linear high-frequency injection:
\begin{equation}
    \hat{x} = D_{\theta}(y) \approx (x * \text{sinc}) + G_{\theta}(y)
\end{equation}
Since this ideal sinc filtering inherently preserves global anatomical context, dense global token mixing—essential for resolving aliasing and non-local similarity in natural images~\cite{chen2023comparative}—offers limited benefits for MRI's locally-dependent details. Consequently, local baselines remain highly competitive, and a mild contextual expansion via our Large-Small Dynamic Gate (LSG) yields modest improvements by aggregating relevant local structural details.

\textbf{Setup.}
We evaluate on IXI~(578 T2 weighted volumes). (https://brain-development.org/ixi-dataset/)
Following~\cite{ji2024deform,yang2025restore}, inputs are generated by retaining the central $6.25\%$ of k-space and zero-filling the remainder, train/val/test contain 40,500/5,828/11,400 slices from 405/59/114 volumes.

\textbf{Dedicated-coil absence Denoising} Dedicated surface coils can substantially boost local sensitivity for carotid vessel wall MRI, but are not always available due to cost and limited deployment.
Following the dedicated-coil absence denoising formulation of~\cite{zeng2025deep}, we study paired restoration from coil-absent acquisitions, which exhibit pronounced spatially varying SNR.
We model the degraded acquisition as a spatially heteroscedastic corruption of a reference image with dedicated carotid coil $x$:
\begin{equation}
y(\mathbf{r}) = g(\mathbf{r})\,x(\mathbf{r}) + \varepsilon(\mathbf{r})
\label{eq:den_forward}
\end{equation}
where $\mathbf{r}\in\Omega$ indexes pixel locations, $g(\mathbf{r})$ captures spatial sensitivity loss due to missing dedicated coils (e.g., an effective sensitivity ratio), and $\varepsilon(\mathbf{r})$ denotes a spatially varying noise scale that increases as sensitivity decreases.

Compared to accelerated reconstruction and super resolution, Eq.~\eqref{eq:den_forward} exhibits strong spatial non-uniformity, motivating token-mixing mechanisms that can aggregate information across distant regions to better infer spatially varying reliability.

\textbf{Setup.}
We collected Simultaneous Non-contrast Angiography and intraPlaque hemorrhage (SNAP)~\cite{wang2013simultaneous} data from 17 volunteers and retrospectively removed the dedicated-coil contribution to obtain paired degraded/reference images, following~\cite{zeng2025deep}.
In total, we obtain 6,290 axial 2D slices.
We perform 17-fold leave-one-volunteer-out evaluation.
All data were acquired on a Siemens Prisma-fit 3.0T system (ages 24--93; 4 female / 13 male).

\subsection{Backbone and Token-Mixing Variants}
\label{sec:backbone}
We establish a minimal gated CNN as our local baseline to benchmark against complex global architectures under aligned protocols. To isolate the effect of expanding receptive fields, we instantiate a large-field variant by altering only the token-mixing operator inside the shared block, serving as a controlled bridge along the token-mixing spectrum, See fig.~\ref{fig1}.

\textbf{Minimal Gated CNN Baseline (NAF)}
\label{sec:naf}
For the image-to-image setting, we adopt NAFNet~\cite{chen2022simple} as a strong \emph{minimal} gated CNN baseline
and train it under our MRI protocols. For reconstruction, we use the same NAF-style block to parameterize $D_\theta$
inside the unrolled scheme (Eq.~(1)).

NAFNet replaces explicit nonlinear activations with a lightweight multiplicative gate~\cite{chen2022simple}.
Given $Z\in\mathbb{R}^{H\times W\times 2C}$, split $Z=[Z_1,Z_2]$ along channels, then
$\mathrm{SG}(Z) = Z_1 \odot Z_2$
where $\odot$ denotes element-wise multiplication.

We refer to the resulting block as our \emph{minimal gated block}, and use it as the shared local-mixing baseline.
%

\textbf{Large--Small Dynamic Gated Block (LSG)}
\label{sec:lsg}
To probe an intermediate design between purely local mixing and full global token mixing, we introduce the
Large--Small Dynamic Gated (LSG) block. We adapt the LS Convolution (LSConv)~\cite{wang2025lsnet} token-mixing
operator---originally proposed for efficient vision networks---to image restoration. Following a \emph{See Large, Focus Small}
strategy, LSG uses broad context to parameterize fine-grained local aggregation.

Given input $X\in\mathbb{R}^{H\times W\times C}$, LSConv first generates position-aware dynamic weights $W$ via a large-kernel perception module:
\begin{equation}
W = \mathrm{PW}_{out}\!\big(\mathrm{DW}_{K_L\times K_L}(\mathrm{PW}_{in}(X))\big) \in \mathbb{R}^{H\times W\times (G K_S^2)}
\label{eq:lsg_weights}
\end{equation}
where $\mathrm{DW}_{K_L}$ is a depth-wise convolution with large scope $K_L$, and $G$ denotes channel groups.
$W$ is then reshaped into dynamic kernels $K_i \in \mathbb{R}^{G\times K_S\times K_S}$ at each location $i$ to perform \emph{small-kernel aggregation}.
The output feature $y$ is computed via grouped dynamic convolution:
\begin{equation}
y_{i,c} = \sum_{(u,v)\in\mathcal{R}_{K_S}} K_i(g,u,v) \cdot x_{i+(u,v),c}
\label{eq:lsg_dynagg}
\end{equation}
By predicting kernels $K_i$ from wide context (Eq.~\eqref{eq:lsg_weights}) applied to local neighborhoods (Eq.~\eqref{eq:lsg_dynagg}), LSConv achieves efficient content-adaptive processing.

We build LSG by swapping the local spatial token mixer in the NAF-style block with $\mathrm{LSConv}(\cdot)$, while keeping the same gated design and residual structure. This yields a lightweight, large-field, context-conditioned mixer that avoids dense all-to-all interactions, providing a more granular testbed that bridges local CNNs and global Transformers—for our task-dependent hypothesis.

\section{Results}\label{sec.result}
\begin{table}[!htbp]
\centering
\begin{threeparttable}
    \caption{Quantitative comparison on the fastMRI single coil knee dataset. PSNR and NMSE are computed under a consistent protocol for all methods, while SSIM is reported under two evaluation protocols to ensure fair comparison. SSIM values are formatted as \textit{Slice-wise (Volumetric)}. Methods with missing values under a specific protocol are denoted by a dash (-).}
    \label{tab:recon_sigle}
    \begin{tabular}{l ccc ccc}
    \toprule
    \multirow{2}{*}{\textbf{Method}} & \multicolumn{3}{c}{\textbf{4$\times$ acceleration}} & \multicolumn{3}{c}{\textbf{8$\times$ acceleration}} \\ 
    \cmidrule(lr){2-4} \cmidrule(lr){5-7}
    & PSNR$\uparrow$ & SSIM$\uparrow$\tnote{a} & NMSE$\downarrow$ & PSNR$\uparrow$ & SSIM$\uparrow$\tnote{a} & NMSE$\downarrow$ \\ 
    \midrule
    
    Zero-filled      & 29.49 & 0.6541 (0.7504) & 0.0532 & 26.84 & 0.5500 (0.6392) & 0.0893 \\
    CS~\cite{uecker2015berkeley}         & 29.54 & 0.5736 (-) & 0.0583 & 26.99 & 0.4870 (-) & 0.0903 \\
    U-Net~\cite{ronneberger2015u}      & 31.88 & 0.7142 (-) & 0.0357 & 29.78 & 0.6424 (-) & 0.0511 \\
    KIKI-Net~\cite{eo2018kiki}    & 31.87 & 0.7172 (-) & 0.0353 & 29.27 & 0.6355 (-) & 0.0546 \\
    Kiu-net~\cite{valanarasu2020kiu}     & 32.06 & 0.7228 (-) & 0.0342 & 29.86 & 0.6456 (-) & 0.0497 \\
    SwinIR~\cite{liang2021swinir}      & 32.14 & 0.7213 (-) & 0.0342 & 30.21 & 0.6537 (-) & 0.0476 \\
    D5C5~\cite{schlemper2017deep}     & 32.25 & 0.7256 (-) & 0.0332 & 29.65 & 0.6457 (-) & 0.0512 \\
    OUCR~\cite{guo2021over}        & 32.61 & 0.7354 (-) & 0.0315 & 30.59 & 0.6634 (-) & 0.0443 \\
    ReconFormer~\cite{guo2023reconformer} & {32.73} & {\bf 0.7383} (-) & {0.0310} & 30.89 & 0.6697 (-) & 0.0429 \\
    HUMUSNet~\cite{fabian2022humus}    & 32.37 & 0.7221 (-) & 0.0323 & 31.04 & 0.6722 (-) & 0.0410 \\ 
    FMT-Net~\cite{yi2023frequency}  & 32.56 & 0.7364 (-) & 0.0441 & 31.06 & 0.6661 (-) & 0.0535 \\
    FPS-Former~\cite{meng2025boosting}  & 32.51 & 0.7337 (-) & 0.0316 & 31.03 & 0.6692 (-) & 0.0408 \\
    PDAC~\cite{wang2024progressive}     & {32.73} & \underline{0.7376} (\underline{0.8201}) & {0.0310} & 31.16 & 0.6739 (0.7687) & 0.0407 \\
    MambaMIR~\cite{huang2024mambamir}     & 32.47 & - (0.8091) & 0.0316 & 30.86 & - (0.7385) & 0.0420 \\
    MMR-Mamba~\cite{zou2025mmr}     & 32.59 & - (0.8161) & 0.0312 & 31.05 & - (0.7709) & 0.0412 \\
    DH-Mamba~\cite{meng2025dh}     & {\bf 32.76} & - ({\bf 0.8211}) & {\bf 0.0308} & \underline{31.17} & - (\underline{0.7741}) & {\bf 0.0403} \\
    \midrule
    \rowcolor[HTML]{F2F2F2} 
    NAFRecon\tnote{b} & {\underline{32.74}} & 0.7375 (0.8198) & \underline{0.0309} & {\bf 31.21} & {\bf 0.6763} ({\bf 0.7743}) & {\underline{0.0404}} \\
    \rowcolor[HTML]{F2F2F2} 
    LSGRecon\tnote{b} & 32.51 & 0.7312 (0.8152) & 0.0320 & 31.10 & \underline{0.6742} (0.7687) & 0.0410 \\ 
    \bottomrule 
    \end{tabular}
    \begin{tablenotes}
        \footnotesize
        \item[a] \textbf{SSIM Protocols:} The primary value represents \textit{Slice-wise SSIM}: $\mathrm{SSIM}_c=\frac{1}{N_c}\sum_{s=1}^{N_c}\mathrm{SSIM}(x_{c,s},\hat{x}_{c,s};\max=x_{c,s}^{\max})$, reporting $\frac{1}{C}\sum_{c=1}^{C}\mathrm{SSIM}_c$. The value in parentheses represents \textit{Volumetric SSIM}: $\mathrm{SSIM}_c=\mathrm{SSIM}(X_c,\hat{X}_c;\max=X_c^{\max})$ computed directly on the full volume, reporting $\frac{1}{C}\sum_{c=1}^{C}\mathrm{SSIM}_c$.
        \item[b] NAFRecon: Minimal Baseline, LSGRecon: Large-field Variant
    \end{tablenotes}
\end{threeparttable}
\end{table}
Prior fastMRI single-coil works report SSIM using different aggregation protocols.
\begin{table}[!htbp]
    \centering
    \caption{The quantitative results of $8\times$ accelerated multi-coil MRI reconstruction using our proposed model and recent methods on the fastMRI knee and Stanford 2D datasets.}
    \label{tab:results}
    \begin{tabular}{lcccccc}
        \toprule
        \multirow{2}{*}{\textbf{Method}} & \multicolumn{3}{c}{\textbf{fastMRI knee}} & \multicolumn{3}{c}{\textbf{Stanford2D}} \\
        \cmidrule(lr){2-4} \cmidrule(lr){5-7}
         & PSNR$\uparrow$ & SSIM$\uparrow$ & NMSE$\downarrow$ & PSNR$\uparrow$ & SSIM$\uparrow$ & NMSE$\downarrow$ \\
        \midrule
        Zero-filled & 27.42 & 0.7000 & 0.0704 & 29.24 & 0.7175 & 0.0948 \\
        U-Net~\cite{ronneberger2015u} & 34.18 & 0.8601 & 0.0151 & 33.77 & 0.8953 & 0.0333 \\
        E2E-VarNet~\cite{sriram2020end} & 36.81 & 0.8873 & 0.0090 & 36.48 & 0.9220 & \underline{0.0172} \\
        HUMUSNet~\cite{fabian2022humus} & 36.84 & 0.8879 & 0.0092 & 35.43 & 0.9134 & 0.0219 \\
        PDAC~\cite{wang2024progressive} & \underline{37.12} & \underline{0.8905} & \bf 0.0085 & \underline{36.77} & \underline{0.9247} & \bf 0.0166 \\
        \midrule
    \rowcolor[HTML]{F2F2F2} 
    NAFRecon(Minimal Baseline) & \bf 37.13 & \bf 0.8911 & \bf 0.0085  &\bf 37.10 & \bf 0.9261 & 0.0198   \\
    \rowcolor[HTML]{F2F2F2} 
    LSGRecon(Large-field Variant) & 36.71 & 0.8832 & 0.0094  & 36.64 & 0.9197 & 0.0224  \\ 
        \bottomrule
    \end{tabular}
\end{table}
To avoid mixing incomparable SSIM scales, we report SSIM under
both conventions in Table~\ref{tab:recon_sigle} and re-evaluate shared reference methods (zero-filled, PDAC) under both protocols as calibration anchors. PSNR and NMSE are computed under a consistent pipeline for all methods.

\subsection{Accelerated MRI Reconstruction}
\textbf{Observation 1.}
In acceleration reconstruction with explicit data consistency, the minimal unrolled gated-CNN backbone achieves highly competitive performance. Consequently, introducing large-field mixing—even in the lightweight form of LSG -- can result in a slight performance decrease.

We attribute this to the fact that global coupling is already explicitly imposed by the Fourier operator and repeated data-consistency steps in the unrolled scheme. In this physics-driven regime, the forward model effectively manages long-range dependencies, rendering additional learned global/large-field token mixing within the regularizer less critical.

\begin{table}[!htbp]
    \centering
    \begin{minipage}[t]{0.48\textwidth}
        \centering
        \caption{The quantitative results of $4\times$ super-resolution}
        \label{tab:mrsr}
        \resizebox{\linewidth}{!}{
            \begin{tabular}{lccc}
                \toprule
                \multirow{2}{*}{\textbf{Method}} & \multicolumn{3}{c}{\textbf{IXI}} \\
                \cmidrule(lr){2-4}
                 & PSNR$\uparrow$ & SSIM$\uparrow$ & RMSE$\downarrow$ \\
                \midrule
                SwinMR~\cite{huang2022swin} & 30.93 & 0.9253 & 32.7339 \\
                F-UNet~\cite{sun2025fourier} & 31.26 & 0.9314 & 31.5675 \\
                Restormer~\cite{zamir2022restormer} & 32.03 & 0.9398 & 29.1343 \\
                X-Restormer~\cite{chen2023comparative} & 29.86 & 0.9150 & 35.9677 \\
                MambaIR~\cite{guo2025mambair} & 31.77 & 0.9369 & 29.8372 \\
                Deform-Mamba~\cite{ji2024deform} & 30.60 & 0.8965 & -- \\
                Restore-RWKV~\cite{yang2025restore} & 32.09 & 0.9408 & 28.9713 \\
                \midrule
                \rowcolor[HTML]{F2F2F2}
                NAFNet~\cite{chen2022simple}(Minimal Baseline) & \underline{32.10} & \underline{0.9415} & \underline{28.9409} \\
                \rowcolor[HTML]{F2F2F2} 
                LSGNet(Large-field Variant) & \bf 32.26 & \bf 0.9422 &  \textbf{28.8624} \\
                \bottomrule
            \end{tabular}
        }
    \end{minipage}
    \hfill 
    \begin{minipage}[t]{0.48\textwidth}
        \centering
        \caption{Quantitative metrics for partial coil denoising on carotid MRI dataset}
        \label{tab:mrdenoising}
        \resizebox{\linewidth}{!}{
            \begin{tabular}{lccc}
                \toprule
                \multirow{2}{*}{Method} & \multicolumn{3}{c}{\textbf{SNAP MRI}} \\
                \cmidrule(lr){2-4}
                 & PSNR$\uparrow$ & SSIM$\uparrow$ & NMSE$\downarrow$ \\
                \midrule
                Input & 15.51  & 0.2282 & 0.1575 \\
                U-Net~\cite{ronneberger2015u} & \underline{19.61} & 0.4561 & 0.0705 \\
                SwinIR~\cite{liang2021swinir}  & 18.52 & 0.4323 & 0.0844 \\
                Restormer~\cite{zamir2022restormer} & 19.52 & 0.4536 & 0.0757 \\
                MambaIR~\cite{guo2025mambair} & 18.92 & 0.4382 & 0.0755 \\
                RWKV-Restore~\cite{yang2025restore} & 19.18 & \underline{0.4584} & 0.0857 \\
            Xformer~\cite{zhang2024xformerhybridxshapedtransformer} & \bf 19.65 & \bf 0.4599 & \bf 0.0646 \\
                \midrule
                \rowcolor[HTML]{F2F2F2}
                NAFNet~\cite{chen2022simple}(Minimal Baseline) & 19.48 & 0.4452 & \underline{0.0665} \\
                \rowcolor[HTML]{F2F2F2} 
                LSGNet(Large-field Variant) & 19.56 & 0.4530 & 0.0704 \\
                \bottomrule
            \end{tabular}
        }
    \end{minipage}
\end{table}
\subsection{MRI Super-Resolution}

\textbf{Observation 2.}
For MRI SR, local convolutional backbones remain strong (see Tab.~\ref{tab:mrsr}). While LSG-based contextual conditioning yields modest improvements, dense global interaction does not demonstrate further advantages.

MRI SR via k-space center cropping is essentially a controlled low-pass degradation. This deterministic process preserves much of the global low-frequency anatomy; thus, recovering missing information mainly requires injecting high-frequency details. This specific requirement is often well served by local processing with limited contextual expansion, rendering dense global token mixing less beneficial.

\subsection{Dedicated-coil Absence Denoising}

\textbf{Observation 3.}
For denoising under pronounced spatial heteroscedasticity, Xformer achieves the best overall performance (Tab.~\ref{tab:mrdenoising}).

Given the limited SNAP cohort, performance gaps require cautious interpretation. However, Xformer's leading performance aligns with both Eq.~\eqref{eq:den_forward} and broader restoration benchmarks: global models are advantageous when corruption is strongly spatially non-uniform. Because noise and effective sensitivity vary substantially here, aggregating distant information helps infer varying reliability and restore corrupted structures. Together, these results support global token mixing as a strong choice for heteroscedastic denoising.

\section{Discussion}
Our experiments show that the utility of global token mixing in MRI restoration is task-dependent, governed by physics-imposed global coupling and the spatial non-uniformity of degradation. In accelerated reconstruction with explicit data consistency, a minimal gated CNN regularizer is already competitive; extra global mixing yields diminishing returns since Fourier encoding and repeated consistency updates propagate information globally. For super-resolution via k-space center-crop, low-frequency anatomy is preserved; local models remain strong for injecting high-frequency details, and LSG brings modest improvements. In contrast, dedicated-coil absence denoising is strongly heteroscedastic, where token mixing better aggregates distant evidence.

By evaluating distinct token-mixing paradigms, we turn degradation analysis into a lightweight selection cue: start with a strong minimal baseline and add global mixing only when long-range aggregation is clearly needed and not already provided by physics. For this research, we focus on common instantiations; conclusions for alternative pipelines (e.g., image-to-image reconstruction, k-space/hybrid, or physics-constrained SR/denoising) remain to be validated.

    

%
%
%

\clearpage

\bibliographystyle{splncs04}
\bibliography{ref}

\end{document}